\documentclass[aps,prl,tightenlines,twocolumn,superscriptaddress,showkeys,showpacs]{revtex4-2}

\usepackage{ae}
\usepackage{amssymb}
\usepackage{graphicx}
\usepackage{amsmath}
\usepackage{amsbsy}
\usepackage{amsthm}
\usepackage{bbm}
\usepackage{epsfig}
\usepackage{color}

\newcommand{\beq}{\begin{eqnarray}}
\newcommand{\eeq}{\end{eqnarray}}

\begin{document}

\title{Quantum Error Channels in High Energetic Photonic Systems}

\author{B. C. Hiesmayr}
\email{Beatrix.Hiesmayr@univie.ac.at}
\affiliation{Faculty of Physics, University of Vienna, Boltzmanngasse 5, 1090 Vienna, Austria}
\author{W. Krzemień}
\affiliation{High Energy Physics Division, National Centre for Nuclear Research, Andrzeja Soltana 7, Otwock, Swierk, PL-05-400, Poland}
\author{M. Bała}
\affiliation{Department of Complex Systems, National Centre for Nuclear Research, Andrzeja Soltana 7, Otwock, Swierk, PL-05-400, Poland}

\begin{abstract}
In medical applications --such as positron emission tomography (PET)-- $511$keV photons that experience Compton scattering are studied. We present a consistent framework based on error-correction channels to fully describe the quantum information-theoretic content of high energetic photons undergoing Compton scattering, characterized by the Klein-Nishina formula in unoriented matter. In this way, we can predict the expected spatial distribution of two or more, pure or mixed, polarization entangled or unentangled photons. This framework allows us to characterize the accessible and inaccessible information for different parameter ranges. It also answers the question of how to describe successive multi-photon scattering. In addition our formalism provides a complete framework for dealing with single and all multi-partite errors that can occur in the propagation, providing the basis for modeling future dedicated experiments that will then have applications in medicine, such as reducing errors in PET imaging.
\end{abstract}

\maketitle

In the domain of low energetic systems such as e.g. currently intensively studied for quantum computing~\cite{QuantumErrorQC} Kraus representations of error channels are one way to characterize errors. In this paper we extend the error correction theory to the domain of high energetic photons. Typically errors that entangle portions of the system of interest with the environment appear to be non-unitary quantum operations, i.e. one can no longer assume or simplify the system of interest as a pure state, our knowledge of the state of interest is not complete. Generally, one assumes that the mathematical object to cover any available information of a quantum system is given by a semi-positive-definite Hermitian density operator
$\rho=\sum_j p_j |\psi_j\rangle\langle\psi_j|\,,$
representing a statistical mixture of different states $|\psi_j\rangle$ that occur with probabilities $p_j$ obeying in general $\sum_j p_j=1$. However, as is well known the decomposition is not unique. Only if one probability equals one, we have a pure state, i.e. it is classically which state is realized. Generally one assumes that an allowed dynamic of a physical system during some time interval $\Delta t$ transforms  $\rho$ into a semi-positive-definite Hermitian matrix $\rho'$ and this dynamic is identified by completely-positive trace-preserving maps~\cite{Nielson}. Therefore, in the so-called Kraus representation, one can describe a transformation by
\begin{equation}
\rho'=\rho(t+\Delta t)=\sum_l K_l \rho(t) K_l^\dagger
\end{equation}
with the completeness relation $\sum_l K_l^\dagger K_l=\mathbbm{1}$, which follows from the invariance of the trace under cyclic permutations of the operators. One important feature of the Kraus representation is that these operators are not unique, i.e. defining $F_l=\sum_k U_{lk}\; K_k$ it follows $\sum_l K_l \rho(t) K_l^\dagger=\sum_l F_l \rho(t) F_l^\dagger$ with $U$ being a unitary transformation.
\begin{figure*}
    \centering
    (a)\includegraphics[width=0.45\textwidth,keepaspectratio=true]{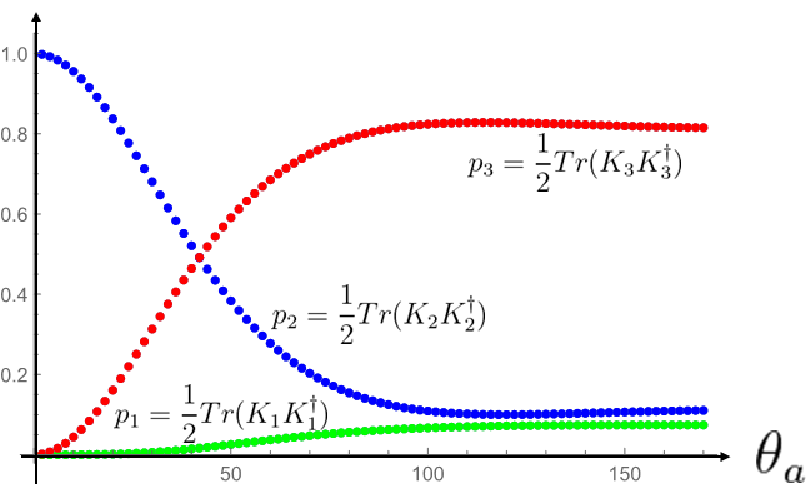}
    (b)\includegraphics[width=0.45\textwidth,keepaspectratio=true]{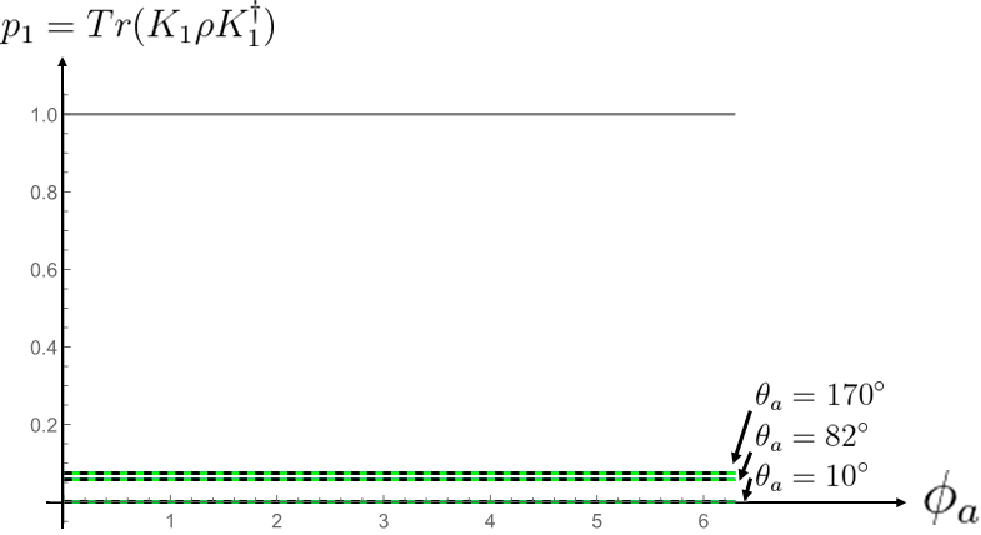}
    (c)\includegraphics[width=0.45\textwidth,keepaspectratio=true]{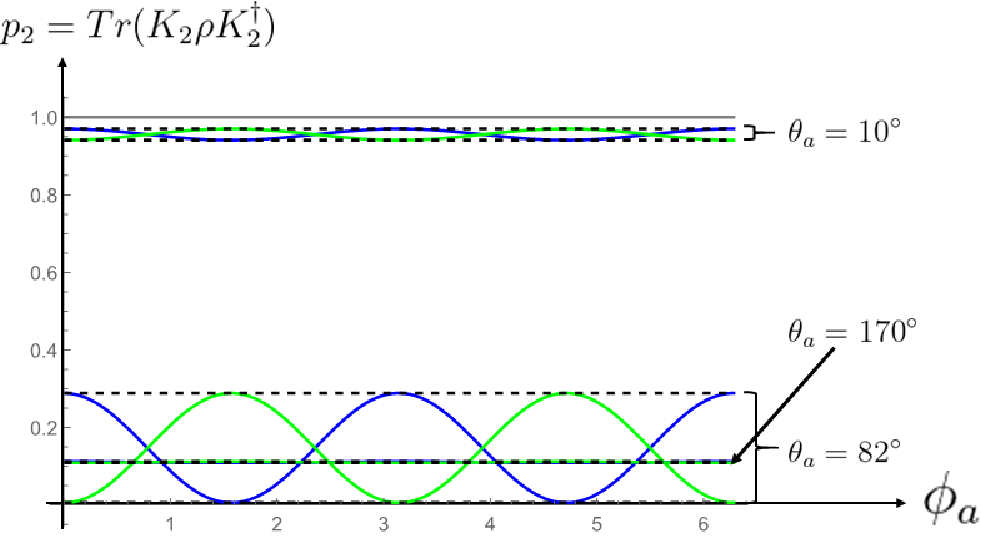}
    (d)\includegraphics[width=0.45\textwidth,keepaspectratio=true]{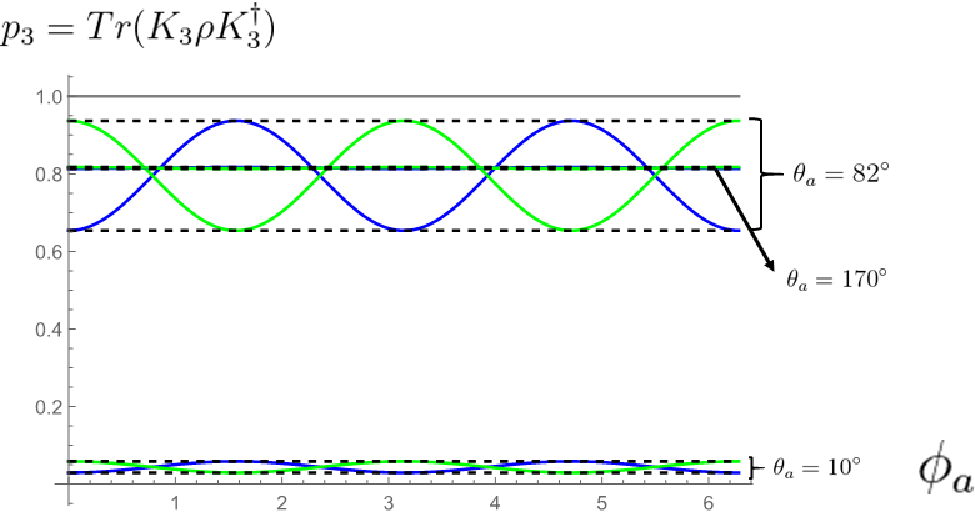}
    \caption{These graphics show the result of the three probabilities $p_1,p_2,p_3=1-p_1-p_2$ deduced from the Kraus operators for (a) unpolarized states and (b)-(d) pure states. In (b)-(d) the blue and green curves corresponds to $|H\rangle$ and $|V\rangle$ polarized states, respectively. The dashed lines corresponds to pure states optimized over unitaries.
    }\label{figqubit}
\end{figure*}

In error correction theory~\cite{Girvin,bennetMixedStateEntPurErrCorr,dynamicalmaps,quantumdephasing} and when working in the interaction picture, the set of Kraus operators $\{K_1,K_2,\dots\}$ is interpreted as the transformation of a state under various errors. So for instance, a set of errors could be a qubit suffering from bit flip $K_1=\sqrt{p_1}\, \sigma_1$ and a phase flip $K_2=\sqrt{p_2}\,\sigma_3$ and therefore the case that no error occurs is given by $K_3=\sqrt{1-p_1-p_2}\, \mathbbm{1}_2$. Typically, one interprets the expectation value
\beq
\langle K_l^\dagger \hat{K_l}\rangle=Tr\left(K_l\;\rho(t)\;K_l^\dagger\right)=p_l
\eeq
as the probability that the $k$th error will occur or has occurred. In error correction theories those probabilities are e.g. exploited to construct logical codewords by the Knill-Laflamme condition~\cite{KnillLaframme}.

In this paper we consider the Compton scattering process of $511$keV photons, initiated by the propagation of photons through a medium described by the Klein-Nishina formula~\cite{Klein-Nishina}, along the theoretic framework described above. In the following, we denote this process as a Compton-Klein-Nishina (CKN) scattering event.

\textbf{Kraus formalism for single CKN scattered photons:} In Ref.~\cite{HiesmayrWitnessing} the authors have presented a pseudo Kraus representation leading to the same result as the Klein-Nishina formula~\cite{Klein-Nishina} when summing over final states. It is called pseudo Kraus representation, because the two Kraus operators do not satisfy the completeness relation. In the following, we present Kraus operators satisfying the completeness relation from which a full description of the CKN scattering events can be deduced, this result is relevant for improving e.g. positron-emission-tomograph (PET) imaging~\cite{toghyani_polarisation-based_2016,jpet1,krzemien_feasibility_2020,jepinpreparation}.

Different to Ref.~\cite{HiesmayrWitnessing} we define the Kraus operators with an overall normalization of the scattering process by
\begin{widetext}
\beq
K_1&=& \frac{\left(\frac{(\sin \theta_s \sin \theta_a \cos (\phi_s-\phi_a)+\cos
   \theta_s \cos \theta_a-1)^2}{-\sin \theta_s \sin \theta_a \cos
   (\phi_s-\phi_a)-\cos \theta_s \cos \theta_a+2}\right)^{3/2}}{\sqrt{2}
   (-\sin \theta_s \sin \theta_a \cos (\phi_s-\phi_a)-\cos \theta_s \cos \theta_a+1)^2} \left(\begin{array}{cc}
 1& 0 \\
 0 & 1 \\
\end{array}
\right)
\eeq
\beq
K_2&=&\left(
\begin{array}{cc}
 \frac{1}{(\cos \theta_s \cos \theta_a-2) \sec (\phi_s-\phi_a)+\sin
   \theta_s \sin \theta_a} & -\frac{i \cos \theta_s \sin (\text{$\phi
   $s}-\phi_a)}{\sin \theta_s \sin \theta_a \cos (\phi_s-\phi_a)+\cos
   \theta_s \cos \theta_a-2} \\
 \frac{i \cos \theta_a \sin (\phi_s-\phi_a)}{\sin \theta_s \sin (\theta
   (a)) \cos (\phi_s-\phi_a)+\cos \theta_s \cos \theta_a-2} & -\frac{\cos
   \theta_s \cos \theta_a \cos (\phi_s-\phi_a)+\sin \theta_s
   \sin \theta_a}{\sin \theta_s \sin \theta_a \cos (\phi_s-\phi_a)+\cos \theta_s \cos \theta_a-2} \\
\end{array}
\right)\;.
\eeq
\end{widetext}
Here the angles $(\theta_s,\phi_s)$ describe the chosen coordinate systems and $(\theta_a,\phi_a)$ the respective change of the propagation direction of the photon undergoing the CKN scattering. The distribution in real space is then given by the probability
\beq
p(\theta_s,\phi_s,\theta_a,\phi_a;\rho)=\sum_{i=1}^2 Tr(K_i\;\rho\; K_i^\dagger)\;,
\eeq
where $\rho$ is the density matrix defining the polarization of the source with respect to the chosen coordinate system $(\theta_s,\phi_s)$. This formula is identical to the scattering cross section of $511$~keV photons undergoing a Compton scattering e.g. in a plastic scintillator or in crystals. Here one assumes that the scattering at such high energies as $511$~keV is described in a good approximation independent of the electron momentum distributions, the orientations of the electron spins and the nuclear spins in the target materials. The qualitative validity of this formula has been shown in several experiments~\cite{jpet1,krzemien_feasibility_2020,jepinpreparation,Watts,makek_single-layer_2020,kozuljevic_study_2021,Russen1,Russen2,McNamara,toghyani_polarisation-based_2016}.

\begin{figure*}
    \centering
    (a)\includegraphics[width=0.45\textwidth,keepaspectratio=true]{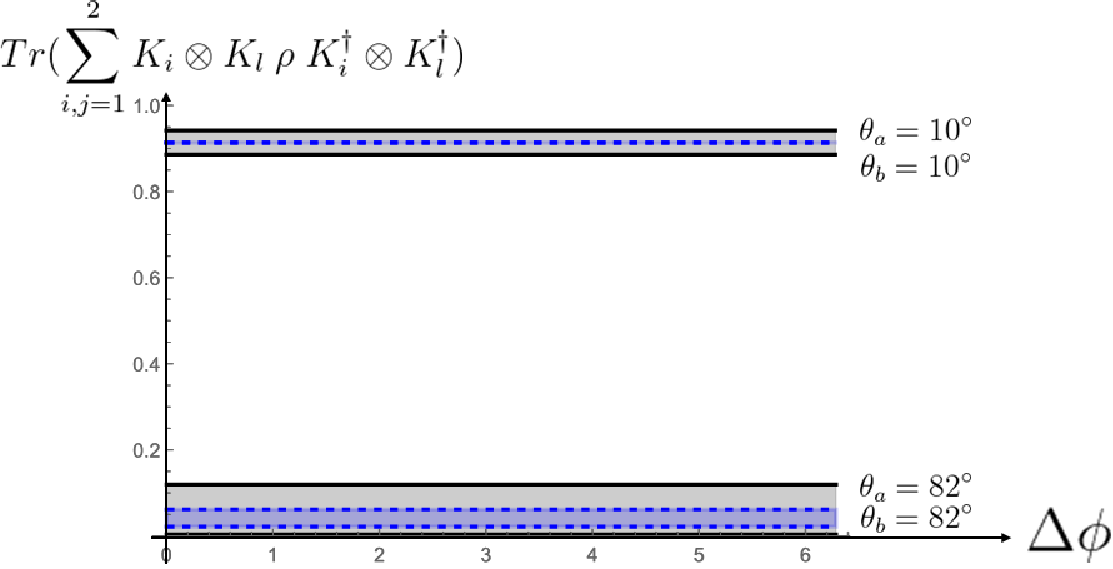}
    (b)\includegraphics[width=0.45\textwidth,keepaspectratio=true]{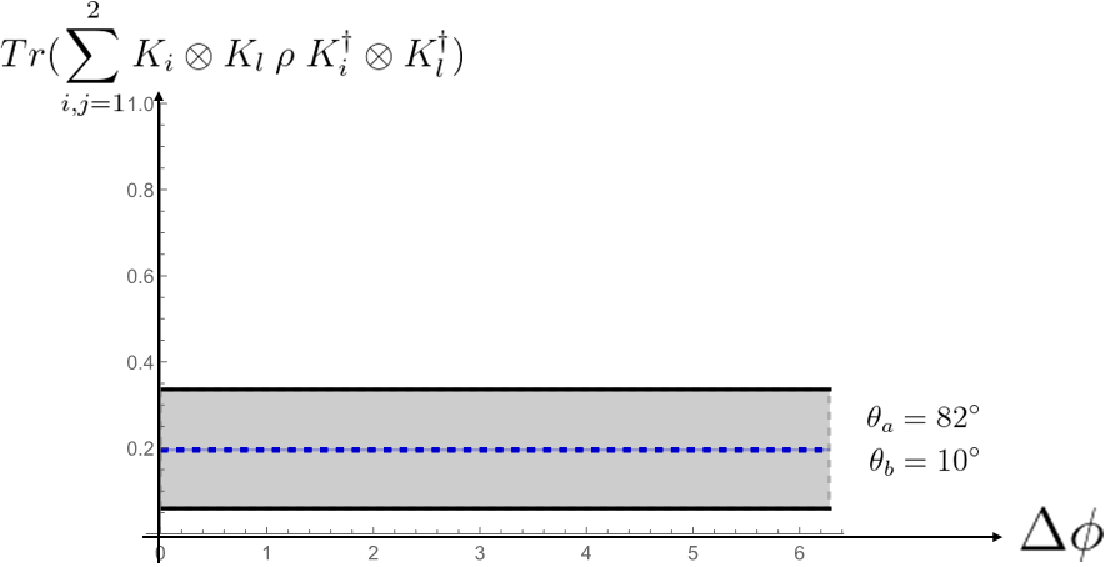}
    \caption{These graphics show the result of the probabilities observed in experiment for pure bipartite states $\rho$ for different scattering scenarios (a) both photons scatter either under $\theta_a=\theta_b=10^\circ$ or $\theta_a=\theta_b=82^\circ$ and (b) one photon under $\theta_a=82^\circ$ and the other one under $\theta_b=10^\circ$ in dependence of $\Delta \phi=\phi_a-\phi_b$ (with $\phi_b=0$). The bold lines bound the grey area which is the region for any initial separable states, whereas the dashed blue lines bound the region for any initial entangled state. The loss of the information is high if the scattering angles are both not small.}\label{figunpol}
\end{figure*}

\textbf{CKN scattering as a quantum channel:} Usually, a system of interest that has no unitary evolution is modelled by adding the environment such that it becomes a closed system and evolves according to the Schrödinger equation. So the question is whether we can find one or more Kraus operators that satisfy the completeness relation.

Indeed, we can find such a completeness relation. A third Kraus operator doing the job, i.e.  $\sum_{l=1}^3 K_l^\dagger K_l=\mathbbm{1}$, can be found (for which we have chosen $\theta_s=0,\phi_s=0$)
\begin{widetext}
\beq
K_3&=&\frac{1}{\sqrt{2}} \sqrt{1-A-\sqrt{(1-A)^2-B^2}} \left(
\begin{array}{cc}
 \frac{\left(1-A-B \cos (2 \phi_a)+\sqrt{(1-A)^2-B^2}\right)}{B} & i\sin (2 \phi_a) \\
 i\sin (2 \phi_a) & -\frac{\left(1-A+B \cos (2 \phi_a)+\sqrt{(1-A)^2-B^2}\right)}{B} \\
\end{array}
\right)
\eeq
\end{widetext}
with
\beq
A&=&\frac{15 \cos \theta_a-6 (\cos (2 \theta_a)+3)+\cos (3 \theta_a)}{8 (\cos \theta_a-2)^3}\nonumber\\
B&=&\frac{\sin ^2\theta_a}{2 (\cos \theta_a-2)^2}\;.
\eeq

In Fig.~\ref{figqubit}~(a) we have plotted the probabilities $p_1,p_2,p_3=1-p_1-p_2$ as a function of the Compton scattering angle $\theta_a$ for a completely unpolarized photon. It is found that the first error, corresponding to $K_1$, increases slowly with the Compton scattering angle, in contrast to the second error, corresponding to $K_2$, which decreases rapidly with increasing Compton scattering angle. For scattering angles larger than $82^\circ$, a maximum loss of about $80\%$ to the environment is reached. In Fig.~\ref{figqubit}(b)-(d), $p_1,p_2,p_3=1-p_1-p_2$ for pure initial states are plotted as a function of $\phi_a$ for three different Compton scattering angles $\theta_a=10^\circ, 82^\circ, 170^\circ$, representing the three relevant ranges. The blue and green oscillating curves are the results for an initial $|H\rangle$ or $|V\rangle$ state. The upper and lower bounds show the unitary optimization over all pure states, i.e., $p_j=\max_{U}Tr(K_j\;U |H\rangle H| U^\dagger K_j)$ and $p_j=\min_{U}Tr(K_j\;U |H\rangle H| U^\dagger K_j)$, respectively. We note that the range of possible values for Compton scattering angles around $82^\circ$ is quite large compared to small or larger values, since in these cases the oscillation term is damped by the visibility of the fringes.

Interestingly, although the information losses to the environment $p_3$ are high (near maximum) at a scattering angle of about $82^\circ$, this is the best region to validate the oscillation of a polarized photon, in contrast to the region of small Compton scattering angles, where the information losses to the environment $p_3$ are small, but the sensitivity to the oscillation is quite small. For large scattering angles, we have large information losses to the environment and low sensitivity to the oscillation.

\textbf{Scattering of bipartite photons and their losses to the environment:} In experiments, one uses sources that produce a photon pair with energy $511$~keV, at which each photon of the pair scatters. It is still open which initial polarization state of a photon pair has to be assumed for typical sources, most experiments assume that the maximally entangled Bell state $|\psi^+\rangle=\frac{1}{\sqrt{2}}\left(|HV\rangle+|VH\rangle\right)$ is the relevant one. However, as we show below also a separable state, i.e.  $\rho_{\textrm{mixed}}=\frac{1}{2}\left(|\psi^+\rangle\langle \psi^+|+|\psi^-\rangle\langle \psi^-|\right)$ with  $|\psi^-\rangle=\frac{1}{\sqrt{2}}\left(|HV\rangle-|VH\rangle\right)$ leads to the same probabilities. Therefore it can be doubted whether the source produces correlations via entanglement.

According to the postulates of quantum theory and assuming the formalism for a single photon CKN scattering is valid
the spatial probability of a photon pair each undergoing a CKN scattering should be given by
\beq\label{doublescattering}
p_\textrm{doubles scattering}(\theta_s,\phi_s,\theta_a,\theta_b,\phi_a,\phi_b,\rho)&=&\\
Tr\left(\sum_{i,j=1}^{2} K_i\otimes K_l\;\rho\; K_i^\dagger\otimes K_l^\dagger\right)\nonumber
\eeq
for any initial two-particle state $\rho$, being either pure or mixed and either being separable or entangled. The spatial distribution has been recorded e.g. for a source of $^{22}\mbox{Na}$ radioisotope emitting positrons, which interact with an electron and then likely form positronium atoms, which subsequently decay into two or three photons~\cite{HiesmayrOrthoPositronium,jpet1}. These photon pairs seem to qualitatively behave according to the above formula~\cite{jepinpreparation}. Researchers currently also develop new sources with different quantum properties~\cite{Brusa}.

Let us emphasize that the probability given by formula~\ref{doublescattering} does not change by choosing for $\rho$ the states $|\psi^{\pm}\rangle$ and $\rho_{\textrm{mixed}}$. This is because in this case the probability $p_\textrm{doubles scattering}$ becomes effectively dependent on $\Delta\phi=\phi_b-\phi_a$. It also reflects the Bose symmetry, and is not sensitive to the sign in the superposition. More details to this can be found in Ref.~\cite{HiesmayrWitnessing}. The loss to the environment is now given by the sum of all contributions including at least one $\mathcal{K}_3$, quantifying in detail what kind of loss in each case is relevant.

\begin{figure}
    \centering
    \includegraphics[width=0.45\textwidth,keepaspectratio=true]{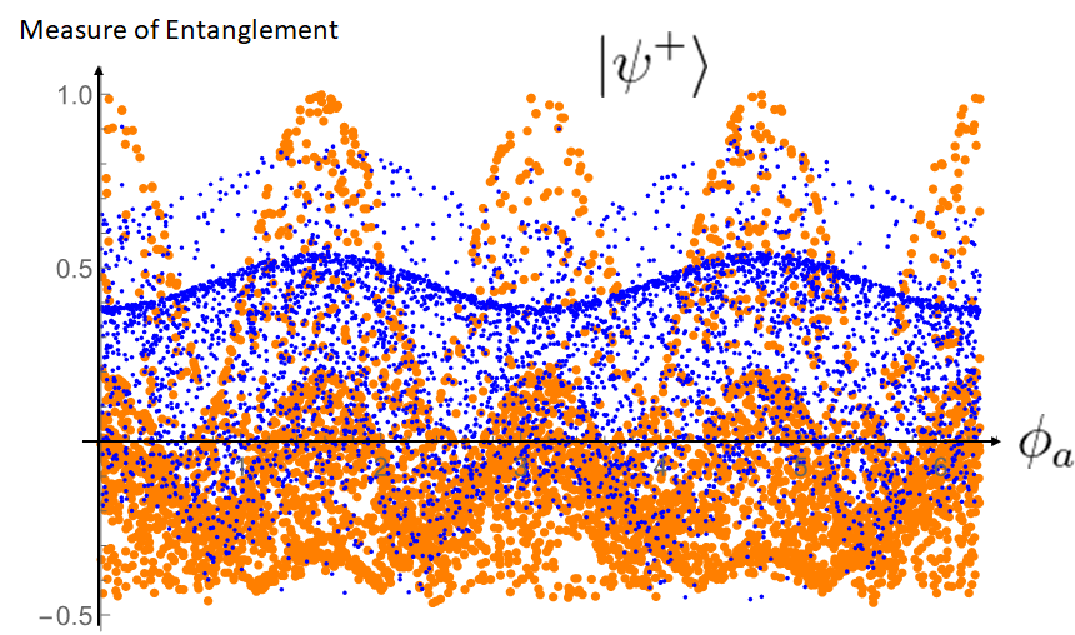}
    \caption{This graphic shows the result of the value of a measure of entanglement (here, the one of Ref.~\cite{quasipure,Wootters}) of $5000$ states generated via
    $\rho=\sum_{i,j=1}^3 K_i\otimes K_j |\psi^+\rangle\langle\psi^+| K_i^\dagger\otimes K_j^\dagger$  (small blue dots) or $\rho_{\textrm{accessible}}=\sum_{i,j=1}^2 K_i\otimes K_j |\psi^+\rangle\langle\psi^+| K_i^\dagger\otimes K_j^\dagger / Tr(\sum_{i,j=1}^2 K_i\otimes K_j |\psi^+\rangle\langle\psi^+| K_i^\dagger\otimes K_j^\dagger)$ (big orange dots) for random generated Compton scattering angles $\theta_a,\theta_b$ and random $\phi_a$ (plotted at the $x$-axis) with $\phi_b=0$. It proves that the CKN double scattering process is an entanglement breaking channel.
    }\label{figent}
\end{figure}

\textbf{One photon in the pair scatters a second time:} Now let us investigate the situation that a photon pair is generated and one of the photons undergoes a CKN scattering a second time. Using the idea that the CKN scattering process introduces an error either $K_1$ or $K_2$ or both, we could think of two different scenarios:, namely either all intermediate states are included $\sum_{c,a,b=1}^{2} K_c.K_a\otimes K_b\;\rho\; K_a^\dagger K_c^\dagger\otimes K_b^\dagger$
or only a particular ``error'' $K_z$ happens $\sum_{c,b=1}^{2} K_c.K_z\otimes K_b\;\rho\; K_z^\dagger K_c^\dagger\otimes K_b^\dagger$.
On the other hand we know that the representation of the Kraus-operators is not unique, i.e. $\sum_l K_l \rho(t) K_l^\dagger=\sum_l F_l \rho(t) F_l^\dagger$ with $F_j=\sum_k U_{jk}\; K_k$. In both discussed cases the $U$ does not cancel out via permutations under the trace, in contrast to the double scattering case~(\ref{doublescattering}). Therefore, we conclude that the CKN scattering process has to be considered a measurement-like process. Thus the photon after the scattering is described by the reduced state and then undergoes a CKN scattering process. This gives probabilities that do not depend on the choice of the Kraus operators
\beq
p_{\textrm{scattered}}=\sum_{c,a,b=1}^{2} Tr\left(K_c Tr_B \left(K_a\otimes K_b\;\rho\; K_a^\dagger\otimes K_b^\dagger\right) K_c^\dagger\right)\;.\nonumber
\eeq
Note, here $Tr_B$ denotes the trace over all degrees of Bob's system, namely the photon that does not scatter a second time. Again the sum over $c$ is necessary, since we assume no orientations in the scattering material. Consequently, this is the experimental result to be expected  if our assumptions above are valid.

\textbf{Errors in the propagation:} Having established a quantum channel to describe fully the CKN scattering process, we also established a theoretical framework to handle potential errors. Here, we have now different possibilities such as an error occurring to one photon or simultaneously both photons and so on. If the error is of a locally unitary form, i.e. $U_1\otimes U_2$, no difference is expected for the probability $p_\textrm{doubles scattering}$ nor $p_{\textrm{scattered}}$ due to the permutation symmetry under the trace. Thus only a global unitary error or a non-unitary interaction in one subsystem introduces deviations. Whether such experimental settings with high enough statistics are achievable must be clarified by further studies out of the scope of this contribution.

\textbf{Property of the CKN error channel:} In Fig.~\ref{figent} we show that the channel describing the double scattering of the pair is entanglement breaking~\cite{EntanglementBreaking} if the initial state is assumed to be a Bell state. On the other hand starting with the separable mixed state $\rho_{\textrm{mixed}}$ for all values, we found no revival of entanglement. Counter-intuitively, the information in principle accessible via experiments (big orange dots) shows higher values of entanglement than the corresponding information for the state including the environment (small blue dots). This is  due to the entanglement monogamy~\cite{monogamy}, i.e. the fundamental property that entanglement cannot be freely shared between arbitrarily many parties.

\textbf{Summary and Outlook:} For a long time it was unclear how to treat Compton scattering at high energies from the point of view of quantum information theory: Is the scattering a measurement process? What information about the polarization is revealed? What is the state of the photon after another scattering process? How do errors propagate? We have answered these questions by presenting a consistent framework that treats scattering as a quantum error channel. This theoretical framework will be the starting point for the development of a Monte-Carlo simulator that will finally allow comparison with experimental data. This in turn will provide the basis for improving, for example, PET imaging based on two or three initial photon events by recording all scattering events.

\textbf{Acknowledgments:} BCH gratefully acknowledges  that this research was funded in whole, or in part, by the Austrian Science Fund (FWF) project P36102. For the purpose of open access, the author has applied a CC BY public copyright licence to any Author Accepted Manuscript version arising from this submission.


\end{document}